# Graphene structure modification under tritium exposure: $^3$H chemisorption dominates over defect formation by $\beta^-$ radiation

*Alexandra Becker[1,2]\*†, Genrich Zeller[3], Holger Lippold[1], Ismail Eren[1], Lara Rkaya Müller[4], Paul Chekhonin[5], Agnieszka Beata Kuc[6], Magnus Schlösser[3], Cornelius Fischer[1,2]*

[1]Helmholtz-Zentrum Dresden-Rossendorf, Institute of Resource Ecology, Department of Reactive Transport, Permoserstraße 15, 04318 Leipzig, Germany

[2]Leipzig University, Faculty of Chemistry and Mineralogy, Johannisallee 29, 04103 Leipzig, Germany

[3]Karlsruhe Institute of Technology, Institute for Astroparticle Physics, Tritium Laboratory Karlsruhe, Hermann-von-Helmholtz-Platz 1, 76344 Eggenstein-Leopoldshafen, Germany

[4]Federal Institute of Technology Zurich, Department of Chemistry and Applied Life Sciences, Vladimir-Prelog-Weg 1-5/10, 8093 Zurich, Switzerland

[5]Helmholtz-Zentrum Dresden-Rossendorf, Institute of Resource Ecology, Department of Structural Materials, Bautzner Landstraße 400, 01328 Dresden, Germany

[6]Helmholtz-Zentrum Dresden-Rossendorf, CASUS, Center for Advanced Systems Understanding, Conrad-Schiedt-Straße 20, 02826 Görlitz, Germany

ABSTRACT. Potential structural modifications of graphene exposed to gaseous tritium are important for membrane-based hydrogen isotope separation. Such modifications cannot be explained by electron irradiation alone. Instead, tritiation, caused by the tritium radicals remaining after the decay, is the primary effect causing the modification of the graphene surface, as confirmed by confocal Raman spectroscopy. The effect of the interaction of tritium atoms with the graphene surface exceeds that of electron irradiation at the average energy of the beta particles (5.7 keV). Compared to previously investigated high electron doses in the absence of tritium, remarkably low concentrations of tritium already induce a significant amount of $sp^3$- and vacancy-type defects at short exposure times. Our findings are supported by molecular dynamics simulations of graphene bombardment with tritium atoms. As a consequence, tritium saturation of graphene may alter its permeability for hydrogen isotopes, thus affecting potential applications.

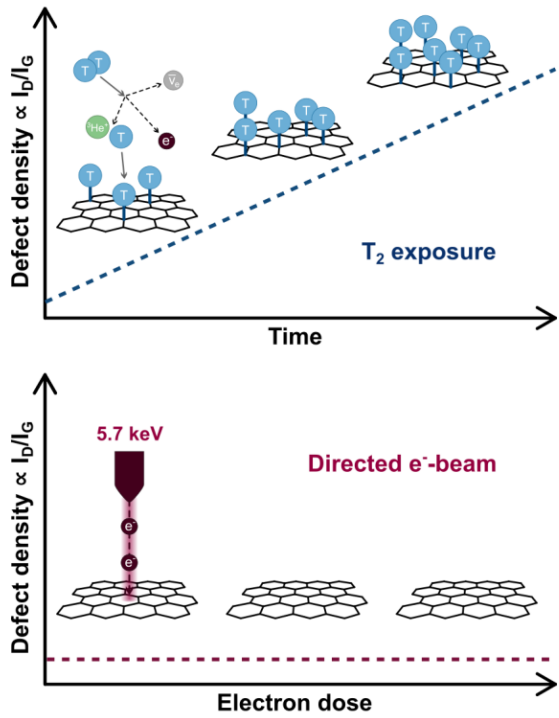

Graphical abstract

1. INTRODUCTION

1.1. BACKGROUND. Nuclear fusion is deemed a promising perspective to help cope with the increasing demand for energy from sources with a low carbon footprint.[1] Tritium, the radioactive isotope of hydrogen, is supposed to be used as fuel in combination with deuterium. Furthermore, tritium is widely used as a tracer and in radiopharmaceutical synthesis.[2–5] On the other hand, it is a matter of concern as it is produced in fission reactors and can leak into the environment.[6–8] Especially for the commercial realization of nuclear fusion, the development of efficient tritium processing technologies is required. Among them, highly selective tritium separation methods need to be explored.

A multitude of methods exist to separate and purify hydrogen from multicomponent mixtures, e.g., electrochemical methods.[9,10] For hydrogen isotope separation, established methods are, e.g., the Girdler sulfide process and cryogenic distillation.[11,12] Recently proposed methods, specifically for gaseous isotopologue mixtures, include selective adsorption on nanoporous materials with satisfactory results for hydrogen/deuterium separation, but also selective diffusion/permeation through proton-conducting polymer membranes proved to be promising.[13–15] For these, graphene was additionally used to increase the separation efficiency.[16]

Graphene, first synthesized and identified in 2004, is a two-dimensional carbon monolayer with a hexagonal structure, featuring extraordinary properties.[17,18] These include very high electron mobility, a high optical transparency and high mechanical stability.[19–21] Furthermore, it is generally impermeable for gas molecules.[22] The characteristics of graphene can be modified, e.g., via doping or through the 'engineering' of defects.[23] It is also possible to modify the structure by hydrogenation leading to graphane structures.[24–27] In a recent study in the field of experimental astroparticle physics, Zeller et al. demonstrated the possibility of tritium adsorption on graphene.[28] Tritiated graphene is also considered for use in the PTOLEMY experiment, in which it serves as a target for capturing relic neutrinos.[29]

The interaction between tritium and graphene leads to the question of whether the graphene structure is modified and what kind of modification might be observed. In view of possible applications for isotope separation, structural changes could impact the selectivity after prolonged and repeated tritium exposure. In a first experiment, Xue et al. were able to separate tritium from water using proton exchange membrane water electrolysis (PEMWE), but attributed the high selectivity to the used catalyst, not to the graphene.[30] Tritiated water has not been found to change the graphene structure.[31] However, for gaseous tritium, with its high activity concentration and insignificantly moderated beta particles, the impact of exposure is likely to be more striking.

1.2. DEFECTS IN GRAPHENE. The occurrence of defects in graphene structures is intensively studied, cf. Table S1 (Supplementary Material) for an overview of the most commonly observed defects. Some defects can be introduced during the growth of the lattice, such as grain boundary defects.[32] The structure can also be altered by chemical modifications, such as doping as well as hydrogenation.[23,27] It has been found that vacancy defects are able to migrate through the graphene structure.[33] Additionally, it has been shown that the graphene structure is altered as a result of electron irradiation.[34]

The most abundant defect type is the Stone-Wales (SW) defect, in which the bond between two carbon atoms has been rotated by 90°, thus forming two 5-rings and two 7-rings.[35] The SW defect has been theorized to aid in hydrogen isotope separation.[36]

1.3. IDENTIFICATION OF DEFECT STRUCTURES IN GRAPHENE. For analyzing the quality of graphene, one of the most common methods is Raman spectroscopy, due to the characteristic spectrum consisting of the so-called G- and 2D-bands, at ~1580 cm$^{-1}$ and ~2700 cm$^{-1}$, respectively.[37] With an increase in defects, the D-band emerges.[38] Thus, the intensity ratio between the D- and the G-band can be used to evaluate the defect evolution in graphene.[39] Additionally, the method is non-destructive, allowing further analysis of the samples.

Another method is transmission electron microscopy (TEM), which can visualize defects while generating them by electron bombardment during the measurement.[40] It can be used to image defects at the atomic level, showing the characteristic structures of, e.g., Stone-Wales defects or vacancies.[41] Scanning tunneling microscopy (STM) is also a suitable method for graphene imaging.[42] X-ray photoelectron spectroscopy (XPS) has been used to measure structural changes, e.g., induced by dopants.[43,44] These methods were not used in this work, because the handling of tritium-containing samples requires specific equipment.

1.4. EXPERIMENTAL APPORACH. As it was first demonstrated by Zeller et al., the exposure of graphene to tritium leads to chemical adsorption of tritium as well as to vacancy-type defects.[28] However, it remained unclear how the vacancy defects are created. Changes in the graphene structure can potentially be caused by the beta electrons or by ions formed by the decay. The respective impact of these two mechanisms on structure changes of graphene is unknown and needs to be studied. On the one hand, we irradiated graphene samples in a SEM with 5.7 keV electrons, which correspond to the average kinetic energy of the tritium decay electrons. Furthermore, we investigated if exposure to tritium gas at comparable electron doses impacts the graphene structure in the same fashion as electrons do.

To provide a more in-depth perspective into the energetics and kinetics of the formation of different defect types, we employed computational studies. These allowed us to predict the defect behavior under various conditions and to characterize the interaction of tritium with graphene as a function of both energy and defect presences.

2. MATERIALS AND METHODS

2.1. GRAPHENE SAMPLES. Irradiation and measurements were conducted with 5 mm × 5 mm monolayer graphene samples on a Si/SiO$_2$ substrate (ACS Material LLC, Pasadena USA) (Figure S2). The graphene was placed on the substrate by the manufacturer using their proprietary *Trivial Transfer Graphene*.[45] The copper substrate on which the graphene is grown is typically removed by etching. To stabilize the graphene during this process, and to easily transfer it to the new substrate, a coating of poly(methyl methacrylate) (PMMA) is applied and dissolved after transfer.[46]

2.2. ORIENTED IRRADIATION: RASTER SCANNING OF SURFACE USING SEM. Four graphene samples have been electron-irradiated inside a scanning electron microscope (EVO-50, Zeiss, Oberkochen, Germany), equipped with a tungsten filament. The scanning was performed over an area of 2 mm × 2 mm, using a defocused electron beam to ensure uniform irradiation. The acceleration energy was set to 5.7 keV, as this corresponds to the average energy of the electrons emitted by tritium decay. The constant beam current was set to $(10 \pm 3)$ nA. It should be noted that the actual beam current generally depends on the SEM gun setup. Different electron doses were applied, calculated from the beam current and irradiation duration. The samples were irradiated for 1.00 h, 2.02 h, 2.68 h and 4.05 h, respectively. All SEM irradiation conditions are summarized in the supplementary information (Table S2).

2.3. DIRECT EXPOSURE TO GASEOUS TRITIUM. To expose the graphene samples to tritium gas, a gas-tight setup was used to exclude any discharge of radioactive material. This was achieved by using a stainless steel vacuum compartment system (RC TRITEC, Teufen; Switzerland). An exposure cell with a volume of 5 ml was attached to the system (Figure S1). Four samples were treated at durations of 22.89 h, 45.78 h, 68.68 h, and 91.57 h, respectively, at room temperature and a tritium pressure of 100 mbar, amounting to approximately 44 GBq tritium. The doses applied by SEM irradiation correspond roughly to only the number of primary decay electrons, not including secondary effects.

2.4. CONFOCAL RAMAN MICROSCOPY. The Raman measurements were performed at the Tritium Laboratory Karlsruhe (TLK) with a confocal Raman microscope (CRM) that was specifically built for the measurement of toxic or radioactive samples.[47] All measurements were carried out with the same configuration using a 532 nm laser with a power of 120 mW. This corresponds to a power density on the graphene surface of about $3 \times 10^5$ W/cm$^2$. At this power density, the graphene layer is not damaged even after many hours of exposure to the laser beam.

The spectrometer of the CRM was equipped with a 1200 g/mm grating, resulting in a resolution of $(9.6 \pm 0.7)$ cm$^{-1}$. For the spectra presented in this work, no spectral sensitivity calibration of the detection system was performed. Therefore, only qualitative comparisons to literature data are possible, also due to the fact that in most of the literature, no statement about the intensity calibration of the used Raman system is made. The data within this work can, however, be compared quantitatively, since only relative changes are discussed, which are not affected by the intensity calibration.

For the used graphene samples, the measured Raman signal was about 300 times weaker than for previously measured samples provided by Graphenea, used by Zeller et al.[28] In comparative measurements against samples by other suppliers, it was demonstrated that the weak signal can actually be related to the sample quality and not to the specific measurement. This necessitated a considerably increased acquisition time for every single spectrum of 300 s and averaging of multiple consecutive spectra (10 - 20) to obtain an acceptable signal-to-noise ratio.

As a consequence, raster scanning of the graphene samples was impractical, and the data analysis required additional steps, especially in regarding the increased number of cosmic ray signals and background fluorescence in each spectrum. To remove the cosmic rays, we used a modified z-score based algorithm, which was applied two consecutive times.[48] Afterwards, the spectrum was smoothed using the Savitzky Golay filter from the SciPy library.[49–51] The background fluorescence was removed. For this purpose, a baseline estimation was performed using an asymmetric least square smoothing algorithm.[50] The remaining constant background was subtracted. Seven peaks (D-, G-, D'-, and 2D-peak of graphene, $N_2$-, $O_2$-, and Si-3TO-lines) in the spectra were fitted simultaneously with Lorentzian functions.[52]

2.5. HEATING OF EXPOSED TRITIUM SAMPLES. The motivation for the bake-out of the graphene samples was to quantify the activities in terms of the amount of bound radiolytic tritium. Using a tritium compatible oven, activities released from the graphene samples were measured. In short, the exhaust from the oven, mostly $T_2$ and HTO, passes through an oxidizing CuO-wire bed with a jacket heater operated at 450°C, and then through a water bubbler, where all tritiated species are retained. The content of the water bubbler is then analyzed by liquid scintillation counting (LSC) to determine the total activity released during the sample heating. The completeness of recovery was recently demonstrated.[53] All four tritiated samples were heated at 300°C for 3.5 h in a stream of argon gas. Time and temperature were chosen for comparability to the investigations by Zeller et al.[28]

2.6. THEORETICAL APPROACH TO THE INTERACTION BETWEEN GRAPHENE AND TRITIUM. To better understand the experimental results and the changes in the graphene structure upon exposure to tritium, we performed molecular mechanics simulations of tritium bombardment of graphene with varying kinetic energies. We simulated the adsorption, transmission (penetration), and reflection rates of tritium atoms on graphene, employing the large-scale Atomic/Molecular Massively Parallel Simulator (LAMMPS).[54] We investigated the tritium bombardment on pristine, Stone-Wales 555-777, and 5-8-5 divacancy defective graphene. We employed molecular dynamics (MD) simulations within a $7 \times 7 \times 1$ supercell of pristine graphene and within a $12 \times 12 \times 1$ supercell of defective graphene. The interactions between carbon atoms were modeled using the Adaptive Intermolecular Reactive Empirical Bond Order (AIREBO) potential.[55] To obtain the adsorption, transmission, and reflection rates for tritium atoms with a certain kinetic energy, from 5 eV to 100 eV (with steps of 5 eV), we placed a grid of tritium atoms 4 Å above the layer of graphene and we ran 1000 MD steps, with 0.1 fs time steps, at 400 K, employing the NVT ensemble with the fixed kinetic energy. Incident energies lower than 5 eV were not considered due to an inconsistency of the AIREBO potential. However, it is expected that when the incident kinetic energy increases, the transmission probability increases.

We created different grids of tritium atoms above pristine and defective graphene models, with 995, 2297, and 1221 tritium atoms for pristine, Stone-Wales 555-777, and 5-8-5 divacancy defective graphene, respectively. Each grid point shows the initial position of the tritium atoms with respect to the graphene layer and was calculated as a separate MD trajectory. The adsorption, reflection, and transmission of tritium atoms were classified according to the final distance of the atoms from graphene.

Additionally, we employed metadynamics simulations to investigate the free energy profile of the formation and transformation of Stone-Wales (SW) defects in a graphene monolayer. These simulations were also conducted using LAMMPS and the AIREBO potential.[54,55] The details of the metadynamics simulations are shown in the Supplementary Material.

3. RESULTS AND DISCUSSION

Pristine graphene has two characteristic Raman bands: the 2D-band at ~2700 cm$^{-1}$ and the G-band at ~1580 cm$^{-1}$, as shown in Figure 1. These bands are associated with phonon modes in the absence of defects in the graphene layer. More specifically, an intensity ratio of $I_{2D}/I_G > 2$ is associated with high-quality defect-free graphene. When defects are present in a graphene layer, additional peaks are observed, the most important being the D-band at ~1340 cm$^{-1}$. In the low-defect regime, the intensity ratio $I_D/I_G$ is directly proportional to the defect density and should be close to zero for defect-free graphene.[56,39] Some pristine graphene materials show a D-peak because of defect structures introduced during synthesis.[57] The $I_D/I_G$ ratio from one of our untreated samples (cf. Figure 1) was determined to be 0.2. As shown in Figure S2, the graphene samples, although untreated, show visible defects, which likely lead to the increased intensity ratio.[58] With the used Raman microscope, there is no way to detect and avoid these defects since it does not have the functionality of a light microscope. Therefore, in this work any $I_D/I_G \leq 0.2$ can be attributed to edge defects and grain borders. In general, a distinction is made between the low-defect regime (stage 1) and the high-defect regime (stage 2) of graphene. Starting from defect-free graphene, the $I_D/I_G$-ratio increases as the defect density increases. At a certain threshold of the defect density, the $I_D/I_G$-ratio reaches a maximum. This defines the transition from stage 1 to stage 2. As the defect density continues to increase, the $I_D/I_G$-ratio as well as the overall intensities of all Raman bands decrease.[39,59] Therefore, when discussing and comparing $I_D/I_G$-ratios and corresponding defect densities, it needs to be considered if the graphene sample is in stage 1 or stage 2. Using additional spectral information, e.g., the width of the G- or 2D-band, this ambiguity can be resolved.[60]

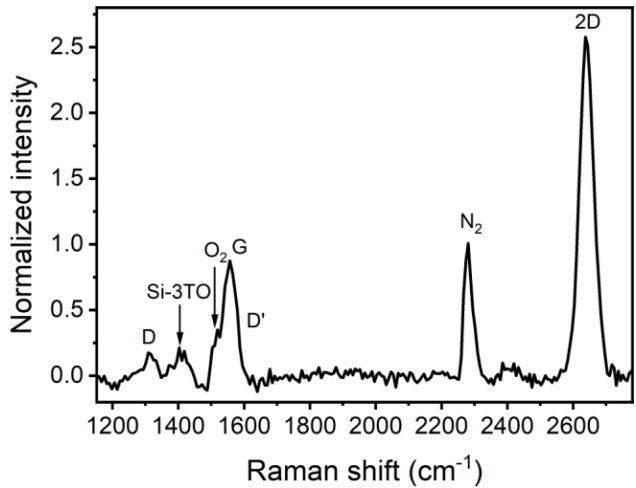

**Figure 1.** Raman spectrum of one of our pristine graphene samples. The spectrum is normalized to its respective G-band. Due to the weak Raman signal from this sample and the increased measurement time, $O_2$ and $N_2$ as well as the 3rd order transverse-optical silicon band (Si-3TO) are visible in the spectrum.

Additional bands in defective graphene are the D'-band at ~1620 cm-1 and the G+D-band at ~3000 cm-1. Of particular interest is the D'-band, because the intensity ratio ID/ID' is sensitive to the different types of defects in graphene. In graphene with sp3-type defects, ID/ID' is ~13, meaning the D'-peak is relatively small. In the presence of vacancy-type defects, ID/ID' is ~7, thus the D'-peak is comparatively larger. Due to the weak Raman signal of the particular graphene samples used for this work, the D'-peak can only be fitted with large uncertainty, and the resulting intensity was close to zero in all cases

3.1. CHANGES IN THE GRAPHENE STRUCTURE CAUSED BY ELECTRON IRRADIATION. Figure 2 shows the Raman spectra of the graphene samples that were exposed to electron irradiation by an SEM. The spectral features show only negligible differences for the increasing electron dose from $0.5 \times 10^{20}$ e⁻/m² to $2 \times 10^{20}$ e⁻/m². Specifically, the normalized intensities and widths of the G-peak and the 2D-peak remain unchanged within the uncertainties of the fit. The spectra show no D-peak, even with increasing electron doses, indicating that no or only few defects have been generated by the SEM irradiation. This also indicates that these measurements were conducted on an edge-defect-free area. Instead of the $I_D/I_G$ peak ratio, the noise-to-G-peak ratio was determined for all measured spectra with an average of 0.12 (Table 1). (The measurement of the sample exposed to an electron dose of $1.0 \times 10^{20}$ e⁻/m² failed because of a defect on the Raman laser.)

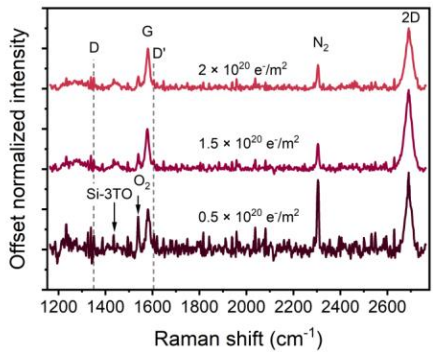

**Figure 2.** Evolution of the averaged Raman spectra with increasing electron doses. The spectra are normalized to their respective G-band amplitudes. Due to the weak Raman signal from these particular samples and the increased measurement times, $O_2$ and $N_2$ as well as the 3rd order transverse-optical silicon band (Si-3TO) are visible in the spectra.

**Table 1.** Noise-to-peak ratios for the SEM irradiated graphene samples. The noise level was determined as the standard deviation for each spectrum at $(1350 \pm 40)$ cm⁻¹ and was divided by the intensity of the measured G-peak.

| Electron dose | Noise-to-peak ratio |
|---|---|
| $0.5 \times 10^{20}$ e⁻/m² | 0.20 |
| $1.5 \times 10^{20}$ e⁻/m² | 0.09 |
| $2.0 \times 10^{20}$ e⁻/m² | 0.08 |

Figure 3 shows an overview plot of results from literature as well as results from this work. In the literature, higher $I_D/I_G$-ratios of up to $I_D/I_G = 1$ were reported at similarly low doses. However, studies reported in the literature employed electron energies above 10 keV. As outlined above, the curves show distinctive trends, first increasing with higher electron doses and then decreasing again. For our work, only low electron doses were of interest. Noticeably, the highest $I_D/I_G$-ratios were observed at 10 keV, with a lower maximum ratio at 30 keV and even lower at 20 keV. Additional literature suggests that changes in the spectra, and thus $I_D/I_G$, may not solely result from amorphization at such low energies, but at least partially from hydrogenation or surface contamination, which can be removed by annealing.[61,62] Tao et al. have reported a small blue shift in the Raman spectra after annealing, indicating a structural modification.[62] It has also been suggested that defects, as visible from Raman spectra, stem from radical compounds formed within the electron beam, leading to chemical etching.[40]

Therefore, we conclude that the oriented beam of 5.7 keV electrons, which mimics the exposure to electrons released by the beta decay of tritium, does not modify the graphene structure.

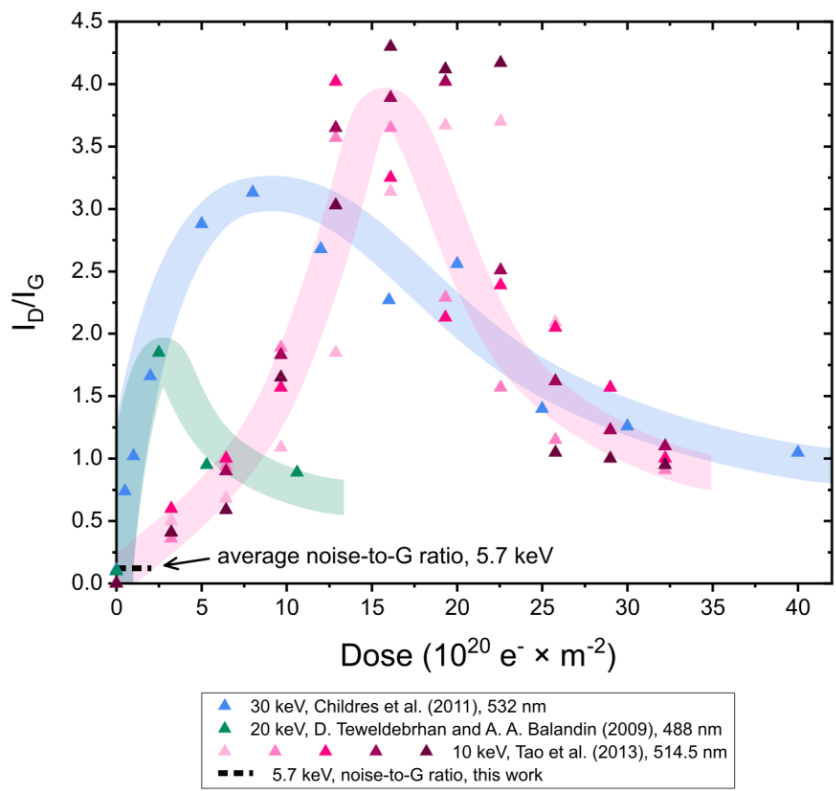

**Figure 3.** Compilation of the $I_D/I_G$ (indicator of defect density) ratio data after SEM irradiation in relation to electron dose. Literature data for 10 keV (pink to magenta), 20 keV (green) and 30 keV (blue) are compared with data from this study (black).[62–64] The stated wavelengths refer to the applied laser beams used. Colored bands have been added to guide the eye.

3.2. EFFECT OF EXPOSURE TO GASEOUS TRITIUM. Four individual graphene samples were exposed to tritium gas at a constant pressure (100 mbar) in a constant volume (5 ml). Only the exposure time was varied. The corresponding Raman spectra of the graphene samples after each exposure are shown in Figure 4A. In general, it can be observed that the D-band intensity increases with longer exposure times, while the 2D-band intensity decreases. Both observations indicate an increasing degradation of the graphene layer. The intensity ratio $I_D/I_G$, which is directly proportional to the defect density in stage 1, caused by hydrogenation as well as other defects, is shown in Figure 4B.

After 22.9 h of exposure to tritium, the changes in the Raman spectra are minimal compared to pristine graphene. The intensity ratio $I_D/I_G$ is less than 0.25, comparable to many pristine samples in the literature.[28,47] The same is true for the intensity ratio $I_{2D}/I_G$ (~1.7). The $I_D/I_G$-ratio is only slightly higher than the noise-to-peak ratio observed for the SEM irradiated samples (cf. Figure 2). For the sample that was exposed to tritium for 45.8 h, a significant change in the Raman spectra can be observed, namely an increase of the D-peak and a decrease of the 2D-peak, leading to an $I_D/I_G$ of ~1.2 and an $I_{2D}/I_G$ of ~1.1. The samples that were exposed to tritium for 45.8 h and 68.7 h show very little difference in the respective Raman spectra. Both intensity ratios remain unchanged within the uncertainties of the fit. After 91.6 h, the D-peak is further enhanced with an $I_D/I_G$ of ~1.5. However, the intensity of the 2D-peak has not decreased significantly, indicating that the general structure of the graphene layer is still intact (region of low-defect graphene, stage 1).

Although the Raman signal of these graphene samples is relatively weak and the D'-peak cannot be fully resolved, it can be inferred that the relatively small D'-peaks are compatible with a certain amount (few %) of $sp^3$-type defects, i.e. a covalent bond between a tritium atom and the carbon of the graphene layer.[60,65] Small D'-peak intensities in general suggest the predominance of $sp^3$-type defects. In contrast, vacancy-type defects produce a higher D'-peak signal intensity, which would be easier to resolve.

We made an estimate whether the number of electrons that interact with the graphene under tritium exposure is comparable to pure electron irradiation. During the tritium chamber irradiation, the number of primary, back-scattered and secondary electrons passing through the graphene layer was estimated with the help of the CASINO software package for scanning electron microscopy.[66] To facilitate an estimation, the geometry of the tritium chamber was simplified as a 2 mm spacing between two infinitely extended walls of stainless steel and the $SiO_2$ substrate. Furthermore, it was neglected that the electrons from tritium upon decay have an energy distribution, as this cannot be done within the CASINO software.

The resulting estimation predicts that out of 100 electrons generated in the tritium gas, the number of electrons passing through the graphene layer is around 55 (primary and back-scattered electrons), plus an additional 47 secondary electrons originating from the $SiO_2$ substrate. Thus, the number of electrons interacting with the graphene is in the same order of magnitude as in the case of SEM irradiation, which justifies a rough comparison of electron doses calculated from beam current and tritium activity. The exposure durations (Figure 4) were set on this basis.

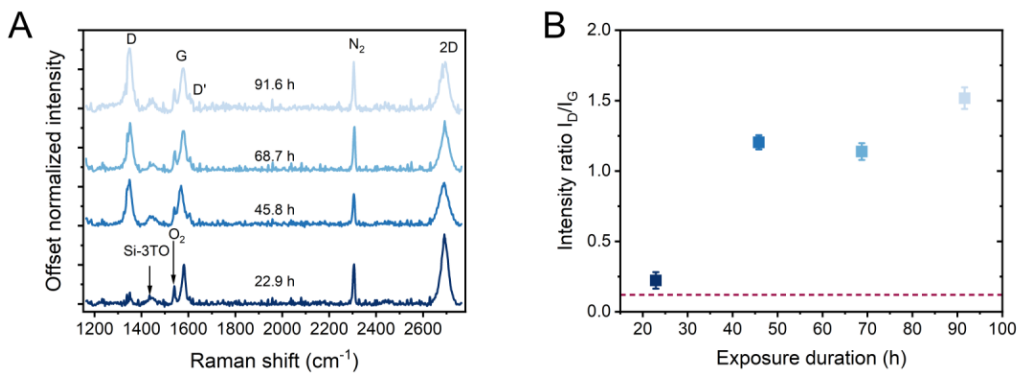

**Figure 4.** A) Evolution of averaged Raman spectra with increasing time of exposure to tritium. B) Evolution of the intensity ratio $I_D/I_G$ as an indicator of changes in the defect density. The red line indicates the average noise-to-G-peak ratio observed after electron irradiation, as shown in Figure 2 and Table 1. The spectra were normalized to the respective G-band amplitudes.

3.3. RELEASED ACTIVITY FROM GRAPHENE SAMPLES. Although it cannot be easily quantified how much of the tritium is adsorbed on the graphene and how much is stored in the SiO$_2$/Si substrates, the study of the released activity after bake-out is a quantitative parameter characterizing the different loading experiments. This also means that this quantity does not depend on the mean free path of different particles under the varying conditions of the cold tritium plasma present in the respective loading setups.

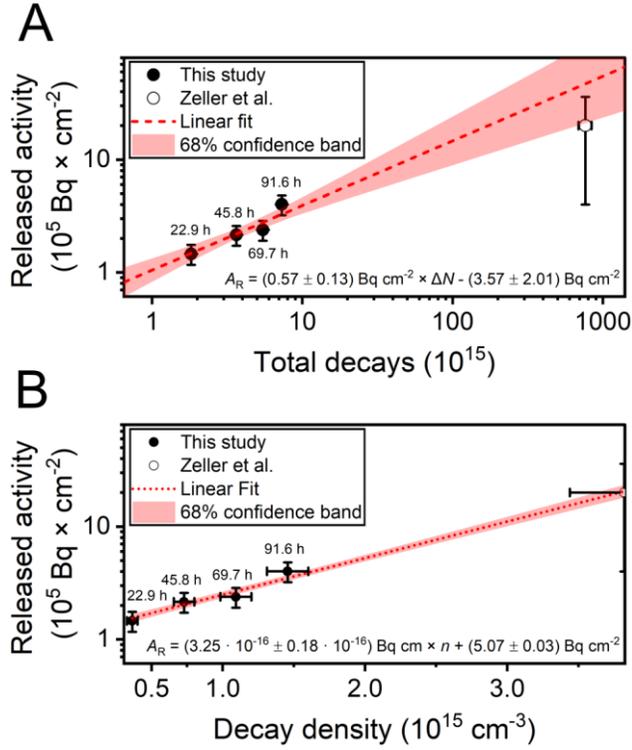

**Figure 5.** Activities measured released after the bake-out of the graphene samples (on Si/SiO$_2$ substrate), corresponding to the amount of radiolytically available tritium. The exposure times of each measured sample are indicated at the data points. Open points indicate data from Zeller et al.[28] A) Plot as a function of the total decays in the chamber during the exposure time. B) Plot as a function of the decay density during the exposure time. The calculations and exact values are given in the supplementary information (Table S3). In the fit equations, $A_R$ denotes the released activity, $\Delta N$ is the total number of decays, and $n$ is the decay density.

Figure 5 shows the released activity as a function of A) the total decays that occurred during the exposure time within the chamber volume in a log-log plot, and B) the decay density in the volume in a log-linear plot. Data points from Zeller et al. are included as open circles. As the area of samples used in the aforementioned study was bigger, a correction factor of 0.25 was applied to the released activity value. For both plots a linear fit was performed as shown.

We want to highlight that compared to the exposure applied by Zeller et al., the total number of decays in this work is two orders of magnitudes smaller (Figure 5A).[28] However, the decay density, based on the volume of the exposure cells, differs only by a factor of about 2.5 (Figure 5B). In the following, we use the released activity $A_R$ as a common parameter as we assume that, given some proportionality factor or offset, $A_R$ is related to the adsorbed activity on the graphene samples and thus also to the defect density in the graphene samples.

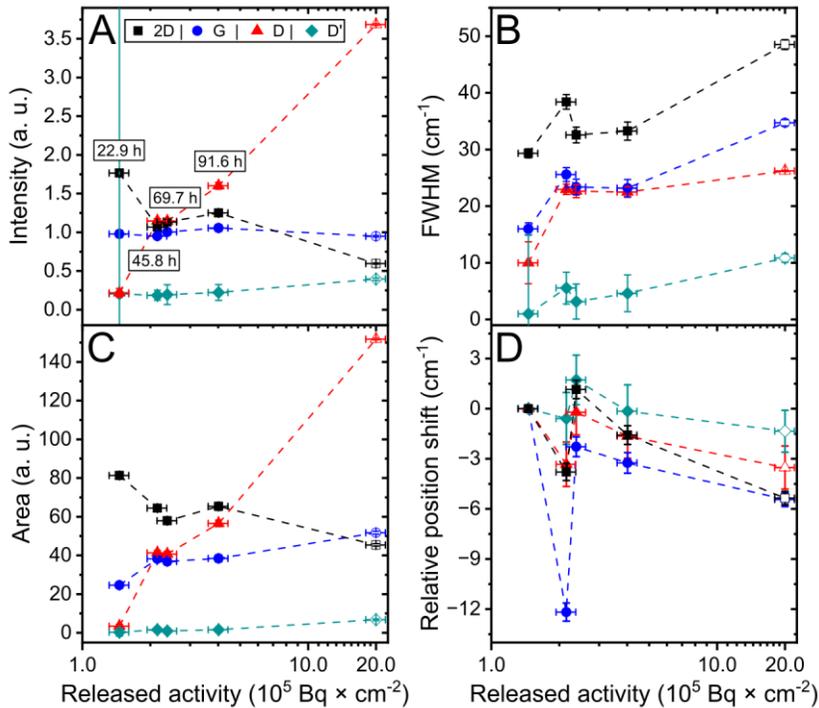

**Figure 6.** Different Raman spectrum peak parameters vs. released activity from graphene on Si/SiO$_2$ samples after heating. The exposure durations are indicated at the data points in chart A.

Open points indicate data from Zeller et al.[28] The dashed lines provide guides to the eye. Note the large error of the D'-peak is caused by the uncertainties of the fit.

$A_R$ is used in Figure 6 to compare the results from this work and the results from Zeller et al. The charts show the peak intensity, peak area, FWHM, and a relative position shift of the main Raman peaks of graphene (2D, G, D, and D'). Note that the data is normalized to the intensity of the G-peak as in Figure 4. Therefore, the intensities shown in Figure 6A can be interpreted relative to the G-peak as $I_D/I_G$, $I_{2D}/I_G$ and $I_{D'}/I_G$.

As it was described previously, the $I_D/I_G$-ratio is directly proportional to the defect density (here estimated by $A_R$) in graphene in the low defect regime (stage 1). This is also observed here: With increasing $A_R$, the D-peak intensity and area (Figure 6C) increases. The $I_D/I_G$-ratio should reach a global maximum at a threshold value of $A_R$, followed by a decrease in the $I_D/I_G$-ratio with further increasing $A_R$. There is some indication that the graphene samples from Zeller et al. (data points at $A_R \cong 20 \times 10^5$ Bq/cm$^2$) correspond to the high-damage density regime of graphene, such as the increased widths of the G- and 2D-peak (Figure 6B). Therefore, the predicted maximum of the D-peak intensity would most likely occur in the region between $A_R \cong 5 \times 10^5$ Bq/cm$^2$ and $A_R \cong 20 \times 10^5$ Bq/cm$^2$.

Noticeably, no significant differences are observed between the trends of the peak intensities (A) and peak areas (C). In both cases, the D-peak value increases with increasing $A_R$, while the 2D-peak value decreases. The G-peak intensity remains approximately 1, as the spectra are normalized to the G-peak maximum. However, the G-peak area increases with increasing $A_R$, which is also reflected in the broadening of the G-peak (B). A significant increase in the D'-peak intensity and area is only visible for $A_R \cong 20 \times 10^5$ Bq/cm$^2$. For $A_R < 5 \times 10^5$ Bq/cm$^2$ it remains constant and small within the fit uncertainties. As discussed previously, the D'-peak is not fully resolved from the noise in this work, also leading to increased error bars of the fit results, most noticeable in Figure 6A. Regarding the peak widths (B), the FWHM of all four peaks increases with increasing $A_R$ except for the sample that was exposed to tritium for 45.8 h during this work. For this sample, all four peaks are significantly broadened. Thus, it appears to be an outlier. A similar effect can be observed when looking at the relative position shifts of the peaks (D). In general, there appears to be a linear anti-correlation between the shift and $A_R$ with the 45.8-h-sample posing an outlier. The G-peak is shifted drastically by $-12$ cm$^{-1}$. It is known that the G-peak position is blue shifted in the case of both electron and hole doping.[67] The line position of $N_2$ (see Figure 4) is stable within $\pm 2$ cm$^{-1}$ across all measurements in this work, indicating that the blue shift is indeed a physical effect.

These results demonstrate a clear correlation between increasing tritium exposure (more total decays, higher decay densities) and the spectral features of the graphene samples. The intensity and area ratios between the D- and G-peak increase, the peaks broaden and there are notable shifts in the peak positions. The results also indicate that with increasing tritium exposure, the released activity $A_R$ is increasing, which suggests that also the amount of adsorbed tritium is increasing. Furthermore, $A_R$ is indeed related to the defect density caused by tritium adsorption. Therefore, a significant amount of the released activity has to originate from the graphene layer itself and not only from the Si/SiO$_2$ substrate.

3.4. THEORETICAL INVESTIGATION OF PRE-EXISTING DEFECTS ON THE INTERACTION BETWEEN GRAPHENE AND TRITIUM. We have investigated by means of simulation what happens when pristine or defective graphene is exposed to tritium bombardment with various initial kinetic energies. The results are summarized in Figure 7, showing the adsorption, reflection, or transmission rates of tritium atoms. In general, increasing the kinetic energy increases the transmission probability and reduces the possibility to form C-T bonds or reflect tritium atoms. In detail, there are two crossing points in Figure 7, which correspond to the crossing between the transmission and either reflection or adsorption. Both points are observed at higher kinetic energies of tritium in case of pristine graphene and are about 5 eV lower in energy for the defective systems, indicating that much less energy is needed to modify the existing defects than to form them from a pristine material.

It is expected that below 5 eV kinetic energy the reflection is lower than adsorption. Additionally to defect formation, the C-T bonds are formed which should be observed in Raman spectra as D'-bands. On the other hand, for higher kinetic energies, the tritium transmission through graphene may result in C atoms knock-out and vacancy formation or defect transformation, in case the tritium atoms are situated above C atoms. The increased formation of defects at higher kinetic energies should be reflected in the Raman spectra as an increased D-band.

The theoretical results show that in the low energy range ($\lesssim$ 20 eV) there is a certain probability ratio between adsorption and reflection of ~1:2, which is rather unaffected by the existing defect type. This is an important aspect, as it suggests that differences in the tritium adsorption efficiency as a function of the defect type are unlikely to occur, and that strongly heterogeneous tritium adsorption behavior is not to be expected. Therefore, we can conclude that the presented results on tritiation are also applicable to graphene substrates that initially show a heterogeneous defect type distribution.

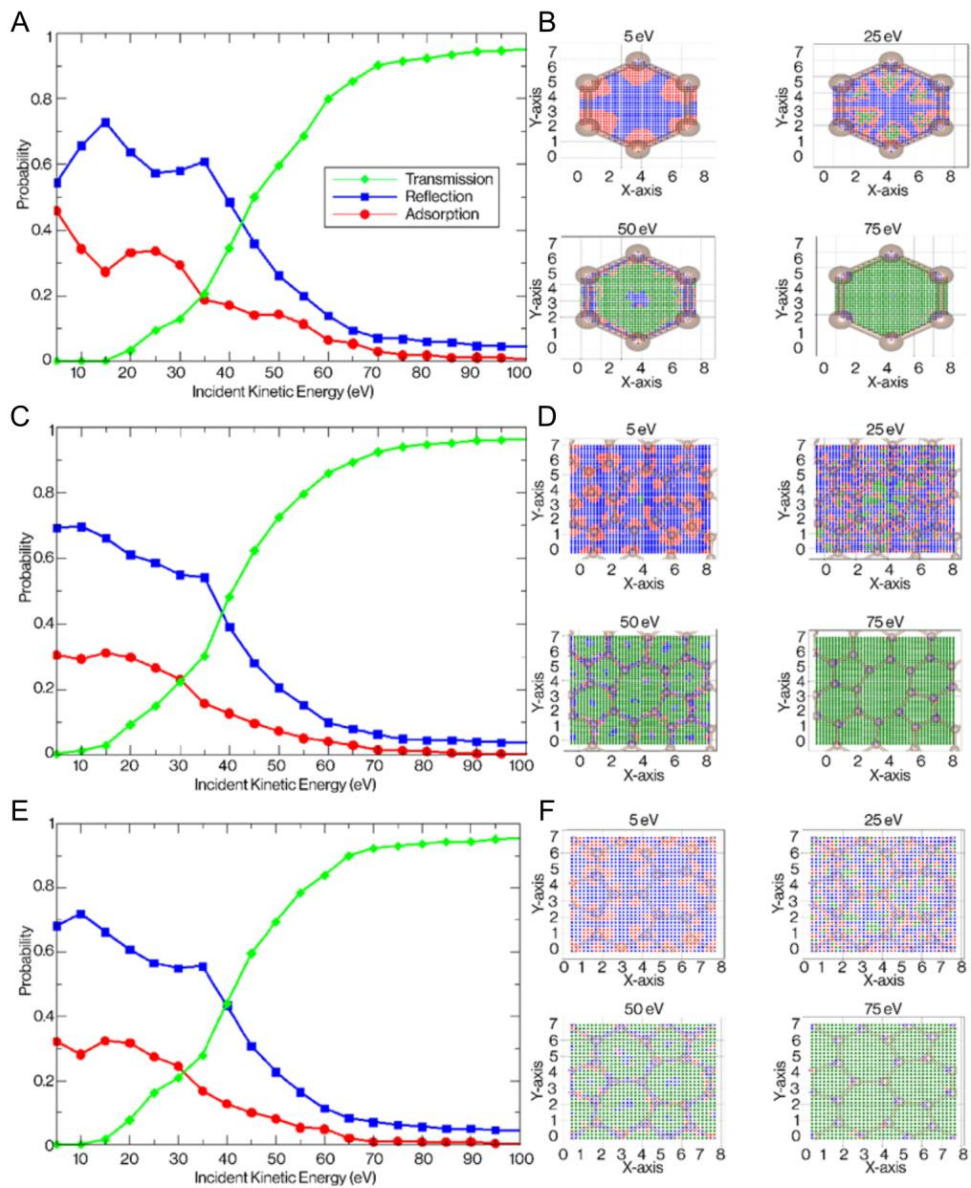

**Figure 7.** Single tritium interaction simulations conducted on pristine graphene, 5-8-5 divacancies, and 555-777 divacancy graphene structures. The probability of single tritium interaction is analyzed as a function of incident kinetic energy, considering transmission, adsorption, and reflection outcomes. Results are presented based on the initial exposure location: (A, B) for pristine graphene, (C, D) for 5-8-5 divacancy graphene, and (E, F) for 555-777 divacancy graphene.

## 4. SUMMARY AND CONCLUSION

We have exposed pristine graphene samples to radioactive tritium gas at the same pressure of 100 mbar with different exposure times. In this gas, primary beta-electrons (keV) as well as secondary particles from ionization and dissociation (low energy electrons, helium and tritium ions and atoms) are present. Our observations show that the interaction between graphene and tritium cannot be explained by electron irradiation alone. The effect of the interaction of the tritium atoms with the graphene surface exceeds electron exposure, as the primary effect observed is tritiation, caused by the tritium atoms remaining after the decay. The common Raman spectra derived $I_D/I_G$-ratio indicates the general defect density, but gives little information about the defect type. Electron irradiation from an electron gun without an interfering gas atmosphere and thus absence of atomic or ionic species showed no significant change in the Raman spectra and a vanishing D-peak. Literature data suggest changes at substantially higher doses. However, the average energy of the beta-electrons (5.7 keV) is too low to cause changes in the graphene structure. It should also be noted that the median energy is even lower.

Due to tritium exposure, the defect density of graphene increases rapidly and at relatively low doses. Compared to previously investigated high electron doses, remarkably low concentrations of tritium already induce significant amounts of $sp^3$- and vacancy-type defects at exposure times as short as 22.9 h. Our tritium bombardment simulations show that there should be a significant C-T bond formation (adsorption and formation of $sp^3$-C atoms) for lower kinetic energies of tritium atoms, while vacancy-type defects are formed for higher kinetic energies (transmission). Previous experiments have shown that the defect density increases with higher tritium exposure. As the $sp^3$-defects are attributed to tritiation, tritium saturation of the graphene may be at least partially possible. This could alter the electrochemical properties of graphene and affect hydrogen transport as well as selectivity in hydrogen isotope separation. Further research is needed to characterize the impact of the induced defects on physicochemical properties of graphene and thus on future applications of graphene membranes.

ASSOCIATED CONTENT

**Supporting Information**. The following file is available free of charge.

Overview of defects in graphene, calculation of electron dose and total decys/decay density, electron irradiation and tritium exposure parameters, sketch of used setup, sample pictures, peak intensity and area ratios vs. FWHM, additional theoretical investigations on defect formation and modification with metadynamics simulations (PDF)

**Research Data.** The research data associated with this publication is available upon request in the RODARE repository under the DOI 10.14278/rodare.3733.

AUTHOR INFORMATION

**Corresponding Author**

*Alexandra Becker - Karlsruhe Institute of Technology, Institute for Astroparticle Physics, Tritium Laboratory Karlsruhe, Hermann-von-Helmholtz-Platz 1, 76344 Eggenstein-Leopoldshafen, Germany; Email: alexandra.becker@kit.edu; https://orcid.org/0009-0004-2903-7242

**Present Addresses**

†Alexandra Becker - Karlsruhe Institute of Technology, Institute for Astroparticle Physics, Tritium Laboratory Karlsruhe, Hermann-von-Helmholtz-Platz 1, 76344 Eggenstein-Leopoldshafen, Germany

**Author Contributions**

Conceptualization: CF, MS, AK; Data curation: AB, GZ, IE; Formal analysis: AB, GZ; Funding acquisition: CF, AK; Methodology: AB, GZ, HL, AK, PC; Project administration: CF; Resources: CF, MS, AK; Software: LRM, IE; Supervision: CF, AK, MS; Visualization: AB, GZ, IE; Writing – original draft: AB, GZ, IE, PC, LRM; Writing – review and editing: CF, MS, AK, HL

All authors have given approval to the final version of the manuscript.

**Notes**

The authors declare no competing financial interest.

ACKNOWLEDGMENT

The authors would like to thank Vanessa Dykas (HZDR) for her support at the graphene irradiation via SEM and Nancy Tuchscherer (KIT) for annealing of the tritium exposed graphene samples.

We gratefully acknowledge funding from the Deutsche Forschungsgemeinschaft (DFG) – Project-ID 443871192 – GRK 2721 "Hydrogen Isotopes $^{1,2,3}$H", to CF and AK.

# Supporting Information

# Graphene structure modification under tritium exposure: $^3$H chemisorption dominates over defect formation by β$^-$ radiation

*Alexandra Becker[1,2]\*, Genrich Zeller[3], Holger Lippold[1], Ismail Eren[1], Lara Rkaya Müller[4], Paul Chekhonin[5], Agnieszka Beata Kuc[6], Magnus Schlösser[3], Cornelius Fischer[1,2]*

[1]Helmholtz-Zentrum Dresden-Rossendorf, Institute of Resource Ecology, Department of Reactive Transport, Permoserstraße 15, 04318 Leipzig, Germany

[2]Leipzig University, Faculty of Chemistry and Mineralogy, Johannisallee 29, 04103 Leipzig, Germany

[3]Karlsruhe Institute of Technology, Institute for Astroparticle Physics, Tritium Laboratory Karlsruhe, Hermann-von-Helmholtz-Platz 1, 76344 Eggenstein-Leopoldshafen, Germany

[4]Federal Institute of Technology Zurich, Department of Chemistry and Applied Life Sciences, Vladimir-Prelog-Weg 1-5/10, 8093 Zurich, Switzerland

[5]Helmholtz-Zentrum Dresden-Rossendorf, Institute of Resource Ecology, Department of Structural Materials, Bautzner Landstraße 400, 01328 Dresden, Germany

[6]Helmholtz-Zentrum Dresden-Rossendorf, CASUS, Center for Advanced Systems Understanding, Conrad-Schiedt-Straße 20, 02826 Görlitz, Germany

* Email: a.becker@hzdr.de

**Table S1.** Overview of selected graphene defect types

| Defect type | Notes | Literature |
|---|---|---|
| Stone-Wales (55-77) | 90° rotation of a C-C-bond within the graphene structure, characterized by two 5-rings and two 7-rings | 1–3 |
| Monovacancy | Produced by ejection of one carbon atom from the graphene structure | 1,4 |
| Divacancy | Formation through ejection of two carbon atoms, multiple structures possible via additional bond rotations. | 1 |
| Adatom | Adsorption of foreign atoms onto the graphene surface | 1 |
| Hydrogenation | Chemical adsorption of hydrogen onto the surface<br><br>Fully hydrogenated 'graphene' is called graphane, in which hydrogen is adsorbed alternately on both graphene sides. | 5–8 |
| Line defects | E.g., grain boundaries, which can occur during growth | 9 |

Calculation of the electron dose from the SEM parameters:

$$\text{electron dose per area} = \frac{\frac{I_{probe}}{e} \times t_{irr}}{A_{irr}} \left[\frac{e^-}{m^2}\right]$$

$I_{probe}$ - probe current

e - elementary charge ($1.6022 \times 10^{-19}$ C)

$t_{irr}$ - irradiation duration

$A_{irr}$ - irradiated area

**Table S2.** Scanning electron microscope irradiation parameters. After irradiation in the SEM, all samples were analyzed with confocal Raman spectroscopy, except sample RGL2, due to a defect on the Raman laser. Sample names were used for differentiation during the measurements and data evaluation

| Sample | e- dose per area (e-/m2) | $A_{irr}$ (m²) | $t_{irr}$ (h) | $I_{probe}$ (nA) | U (keV) |
|---|---|---|---|---|---|
| RGL1 | $0.5 \times 10^{20}$ | $4.506 \times 10^{-6}$ | 1.00 | 10 | 5.7 |
| RGL2 | $1.0 \times 10^{20}$ | $4.541 \times 10^{-6}$ | 2.02 | 10 | 5.7 |
| RGL3 | $1.5 \times 10^{20}$ | $4.163 \times 10^{-6}$ | 2.68 | 10 | 5.7 |
| RGL4 | $2.0 \times 10^{20}$ | $4.553 \times 10^{-6}$ | 4.05 | 10 | 5.7 |

Rasterelektronenmikroskop Graphen Leipzig (RGL)

Calculation of total decays and decay density:

$$\Delta N = N_0 (1-e^{-\lambda t})$$

with

$$N_0 = 2 \cdot \frac{p \cdot V}{k_B \cdot T}$$

and

$$\lambda = 1.787 \cdot 10^{-9}\, 1/s$$

$\Delta N$ - number of decays/decayed tritium atoms

$N_0$ - initial number of tritium atoms

$\lambda$ - decay constant

$t$ - exposure time

$p$ - pressure

$V$ - exposure cell volume

$k_B$ - Boltzmann constant

$T$ – temperature

**Table S3.** Tritium exposure parameters. The sample names were used for differentiation during the measurements and data

| Sample | $\Delta N / V$ (1/L) | $\Delta N$ | $t$ (h) | $N_0$ | $p$ (mbar) | $V$ (L) |
|---|---|---|---|---|---|---|
| TGL1 | $7.2 \times 10^{14}$ | $3.58 \times 10^{15}$ | 22.9 | $2.43 \times 10^{19}$ | 100 | 0.005 |
| TGL2 | $1.4 \times 10^{15}$ | $7.16 \times 10^{15}$ | 45.8 | $2.43 \times 10^{19}$ | 100 | 0.005 |
| TGL3 | $2.1 \times 10^{15}$ | $1.07 \times 10^{16}$ | 68.7 | $2.43 \times 10^{19}$ | 100 | 0.005 |
| TGL4 | $2.9 \times 10^{15}$ | $1.43 \times 10^{16}$ | 91.6 | $2.43 \times 10^{19}$ | 100 | 0.005 |
| TLK | $7.6 \times 10^{15}$ | $1.51 \times 10^{18}$ | 55 | $4.28 \times 10^{21}$ | 440 | 0.2 |

Tritium Graphen Leipzig (TGL), Tritiumlabor Karlsruhe (TLK)

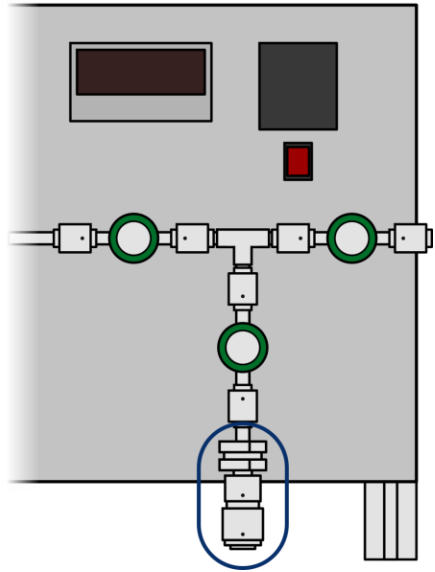

**Figure S1.** Utilized exposure cell attached to the vacuum manifold system (sketch).

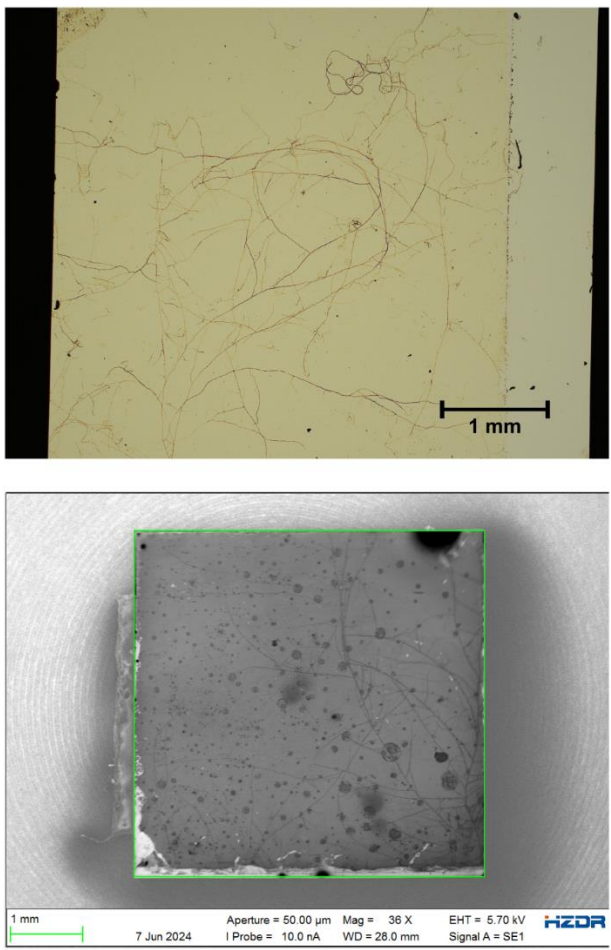

**Figure S2.** Optical microscope (top) and SEM picture (bottom) of two different graphene samples (ACS Material LLC, Pasadena, USA).

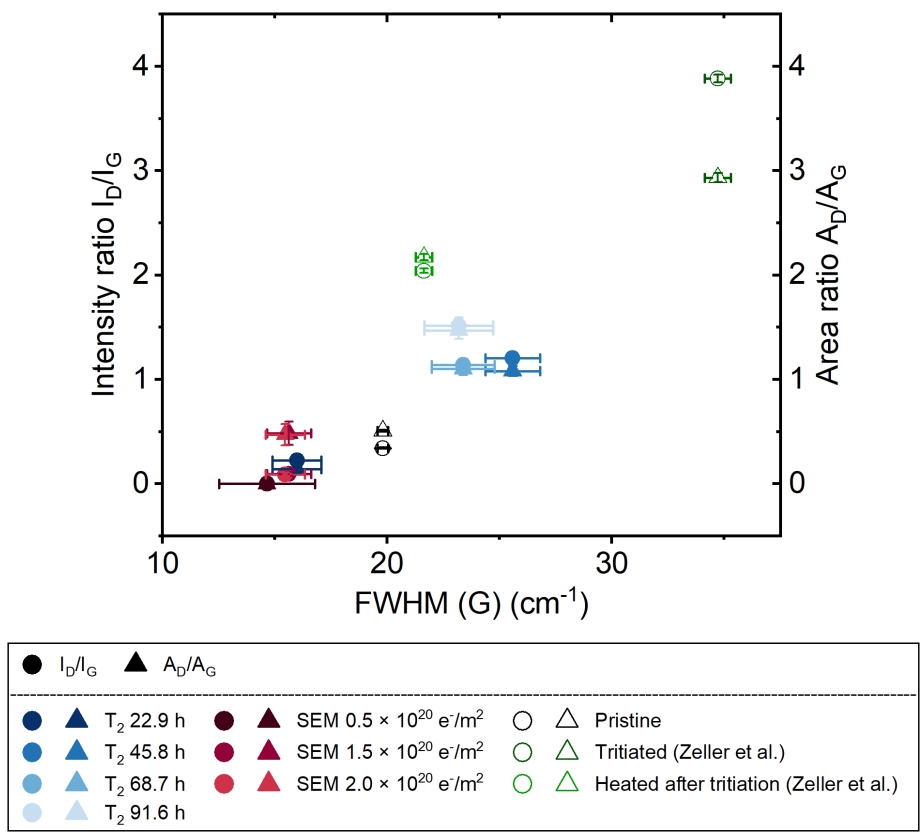

**Figure S3.** Ratios of the peak intensities $I_D/I_G$ (spheres) and peak areas $A_D/A_G$ (triangle) vs. FWHM of the G band for all measured graphene samples. Open data points indicate data from Zeller et al.[10] Plot based on a study by Eckmann et al.[11]

**THEORETICAL INVESTIGATIONS OF DEFECT FORMATION AND DEFECT MODIFICATION.** One of the more commonly observed defects in graphene is the 55-77 Stone-Wales defect, which is characterized by a 90° rotation of a C-C bond. Computational studies can provide valuable insights into the energetics of such a defect formation, allowing for the prediction of defect behavior under various conditions. In a similar fashion, the existing divacancy structures can be further modified to form another defect.

We employed metadynamics simulations to investigate the free energy profiles of the formation and transformation of divacancy defects in a graphene monolayer. These simulations were conducted using LAMMPS[12] with AIREBO potential[13]. The same supercell sizes of graphene models as for the tritium bombardment (as described in the main text) were used. The metadynamics simulations were implemented using the PLUMED library[14,15], version 2.8.1. We defined the collective variables for metadynamics simulations as shown in Figure S4.

The free energy barrier of a defect formation starting from pristine graphene or from an existing defect are shown in Figure S4. Forming a SW defect (55-77) in a pristine graphene requires about 0.60 eV energy barrier. Note that AIREBO underestimates that energy barrier compared to DFT-based methods.[1] However, if a defect already exists, its transformation to another type of defect requires much less energy (only 0.14 eV in the case of a transition from a 555-777 to a 5-8-5 defect or 0.10 eV in case of a 555-777 defect transforming into a 555-6-777 one). In general, it can be concluded that less energy is required to change an existing defect to another one than to create it in pristine graphene. This is also consistent with our findings from the tritium bombardment simulations, shown in the main text (see Figure 7).

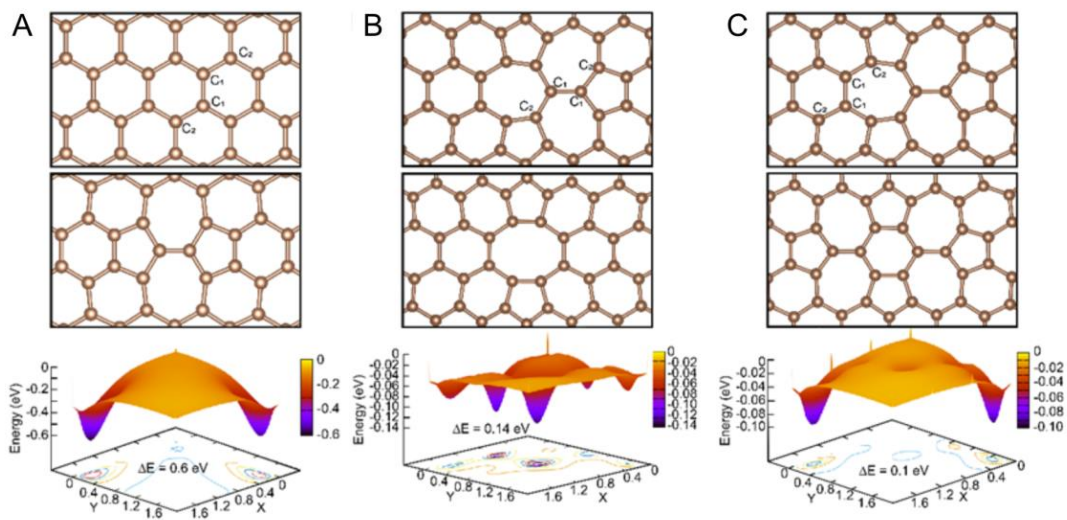

**Figure S4.** Structural description of various graphene formation with and without defect including collective variables to form SW defect (top and middle), and free-energy surface of the defect formation (bottom) in graphene from (A) pristine to 55-77 SW, (B) from 555-777 divacancy to a 5-8-5 divacancy, and (C) from a 555-777 to a 555-6-777 divacancy. The free energy barriers of each transformation are given. X and Y represent the coordination numbers of C atoms used as collective variables in the metadynamics simulation, where their rotation leads to the defect formation or transformation.